\documentclass[a4paper,fleqn,sort&compress]{cas-sc}

\usepackage[numbers]{natbib}
\usepackage{cuted}
\usepackage{soul}
\setstcolor{red}
\usepackage{lineno}
\usepackage{setspace}
\def\tsc#1{\csdef{#1}{\textsc{\lowercase{#1}}\xspace}}
\tsc{WGM}
\tsc{QE}
\tsc{EP}
\tsc{PMS}
\tsc{BEC}
\tsc{DE}

\begin{document}
\let\WriteBookmarks\relax
\def\floatpagepagefraction{1}
\def\textpagefraction{.001}
\shorttitle{Interface properties of CsPbBr$_3$/CsPbI$_3$
perovskite heterostructure}
\shortauthors{J. Kumar et~al.}
\title [mode = title]{Interface properties of CsPbBr$_3$/CsPbI$_3$
perovskite heterostructure for solar cell} 
\renewcommand\linenumberfont{\normalfont\scriptsize\sffamily\color{red}}


\author[1]{Jagadish Kumar}[style=chinese, orcid=0000-0003-1278-7974
] 
\cormark[1]
\ead{jagadish.physics@utkaluniversity.ac.in, Phone:+91-674-2567079}
\credit{Conceptualization, Computation, Analysis, Writing - review editing}
\author[2]{K. P. S. S. Hembram}
\cormark[1]
 \ead{kailashh@alum.iisc.ac.in, Phone: +91-6360905548}
\credit{Conceptualization, Methodology, Analysis, Writing - original draft}

\address[1]{Department of Physics, Utkal University, Vani Vihar, Bhubaneswar 751004, India}
\address[2]{Indian Institute of Science, Bengaluru 560012, India}

\cortext[cor1]{Corresponding authors}
\begin{abstract}
\setstretch{2.0}
We explore the interface properties of perovskite heterostructure CsPbBr$_3$/CsPbI$_3$ through first-principles calculations. The structural interface is formed by the bonding of Cs-Br and Cs-I with bond length of $\sim$4.106 and 3.922 \AA. The upshift of Goldsmith tolerance factor in the range $0.8<t<1$ from $t<0.8$ is revealed for the bi-layer interface, from bulk, reflecting the structural rearrangement from anisotropy to isotropy in confinement. The band gap arises mainly due to the energy difference of I-5p orbital than that of Br-4p at the valence band and Pb-6p at the conduction band. Heavier halide shows the red shift in the absorption spectra, for the pristine monolayer component. For the bilayer geometry, iodine contribution is more observed than that of bromine and the underlying interface properties may be useful for solar cell devices application.

\end{abstract}


\maketitle
\setstretch{2.0}
\section{Introduction}
The use of inorganic perovskites materials for making solar cells is one of the pathways to harvest solar energy \cite{Huang16,Eperon17,Duan19,Xiang19}. They have attracted much attention due to their high power conversion efficiency \cite{Wang18,Shu18,Gao19}, tunable emission wavelength \cite{Protesescu15,Ramasamy16,Shen18}, high photoluminescence (PL) quantum efficiency \cite{Ramasamy16} and excellent optoelectronic properties \cite{Protesescu15,Yantara15,Bwang20,Fadla20}. Because of the superior optical properties of CsPbX$_3$ (X=Cl, Br, I), it has potential application in the area of Light Emitting Diode (LED), photodetectors, laser, dye solar cells etc \cite{Yoon16,Guan20,Yan21,Sidhik17}. The most recent generation of solar cells are mainly consists of polymer based dye-sensitised solar cells (DSSCs) \cite{Carella18,Fagiolari20,Rahman21,Haro21}, organic-inorganic halide perovskite solar cells (PSCs) \cite{Yang21}, organic polymer PSCs \cite{Lu22}, photoelectrochromic devices \cite{Dokouzis20}. But, the metal halide perovskites (MHPs) is a subject of increasing interest because of its marvellous photophysical properties \cite{Parikh22}. Especially, the inorganic hybrid perovskite solar cells such as CsPbBr$_3$ and CsPbI$_3$ are more stable to moisture, heat and light exposure with high power conversion efficiency \cite{Ding21,Kumar21}.

Inorganic halide perovskites structures come with AMX$_3$ (A = alkali ion, M = cation, X = halide). The bonding between two different atoms shows intrinsic interface with dissimilar electron sharing affecting the material properties. For instance, several studies have been carried out with change in alkali metal \cite{Pitriana19,Nam17,Li18}, cation \cite{Shu18,Akkerman17,Hu17,Liang17,Goesten18}, halide \cite{Bian18,Chen18,Haque20,Murtaza11,Yan20} to capture those scenarios with different experiment and theory. Technologically, the desired compositions with defined structures are achieved by simple chemical process, where micro to nano sized samples are formed in the localised scale \cite{Hu17,Liang17,Haque20,Murtaza11,Ghaithan20}.

Understanding of heterostructure and its formation mechanism are important for the manufacture of efficient dye-sensitized and perovskite solar cells. So in this direction, interface of two different materials has been investigated in metallic \cite{Yaniv78,Oliver12}, nitride \cite{Saha11}, oxide \cite{Samal11,Wang20,Chikina21} materials for various technological applications. Bilayer interfaces of many two dimensional materials have been investigated for various purposes \cite{Eshkalak18,Arora21,Kosmider13}. On the solar device geometry aspect, the perovskite material is sandwiched between electron and hole conducting materials, but very few studies have been carried out for the interface of two different perovskite materials \cite{Li10,Yamada02,Tsukada97}. However, the interface at the limiting condition, i.e. a bi-layer, consisting two different perovskite monolayer is hardly investigated. One this aspect, we investigate the structural and electronic properties of CsPbBr$_3$/CsPbI$_3$ at the interface with first-principles calculation.

\section{Computational methods}

Our calculations are based on density functional theory (DFT), which is implemented in the Quantum Espresso (QE) simulation package \cite{Giannozzi09}. Generalized gradient approximation (GGA) was used for exchange correlation energy of electrons \cite{Perdew96}. Ultra-soft pseudopotentials were used to represent the interaction between ionic cores and electrons \cite{Vanderbilt90}. Plane-wave basis with an energy cutoff of 50 Ry and a charge density with a cutoff 300 Ry were used in the calculation. Suitable mesh of k points were used for integration over the irreducible Brillouin zone \cite{Monkhorst76}. Occupation numbers were smeared using the Methfessel-Paxton scheme with broadening of 0.003 Ry \cite{Methfessel89}.

\section{Results}
\begin{figure}
\centering
\includegraphics[height=7.0cm, width=10.5cm]{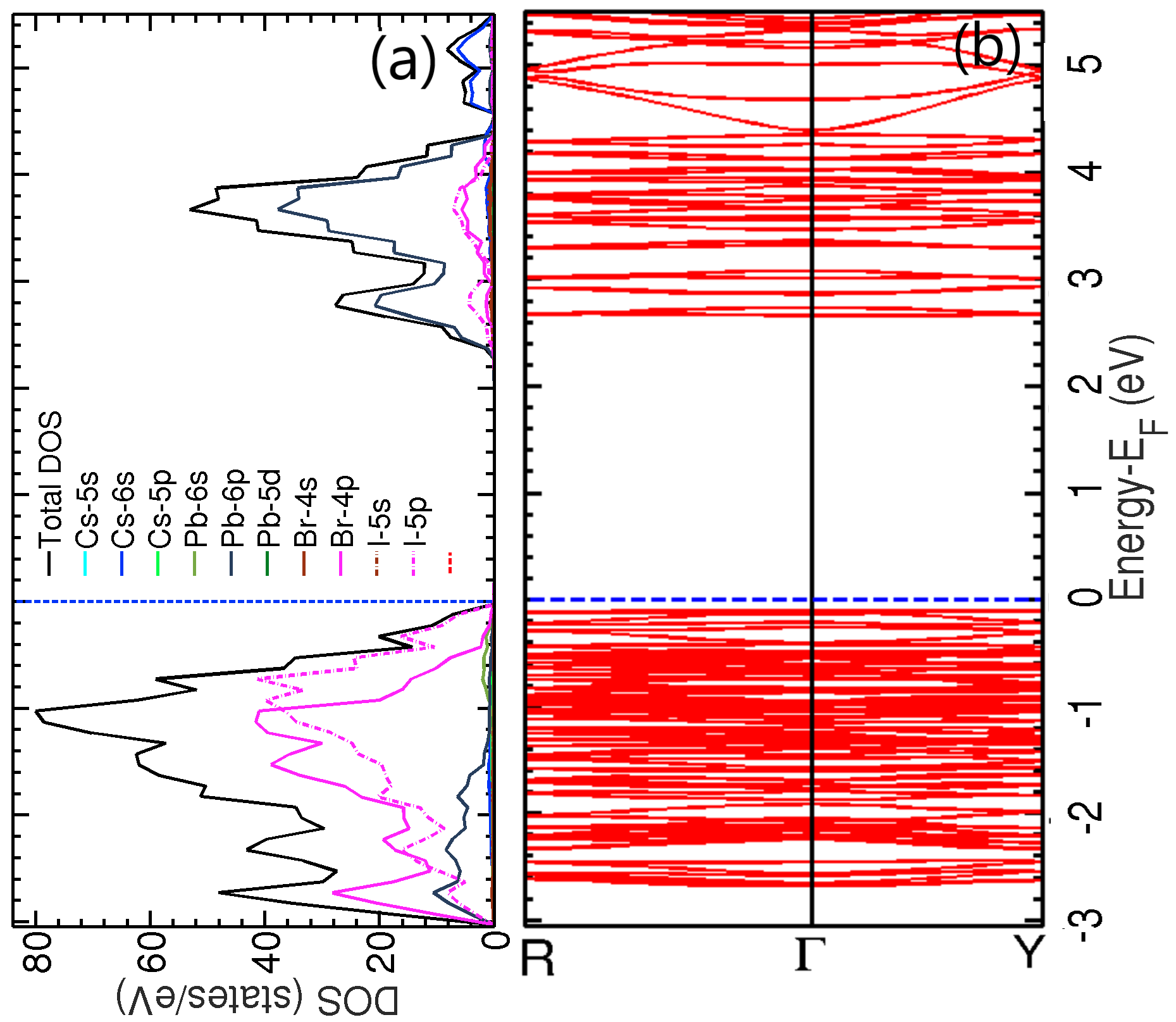}
\includegraphics[height=6.0cm, width=10.8cm]{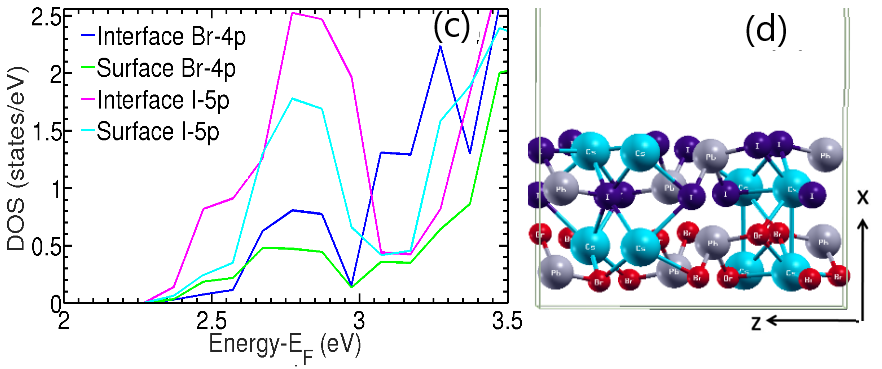}
\caption{(a) Electronic partial density of states, (b) electronic band structure of CsPbBr$_3$/CsPbI$_3$ bi-layer. (c) Orbital contribution of Br-$4p$ and I-$5p$ near the Fermi level of interface and surface atoms. (d) Cross sectional view of CsPbBr$_3$/CsPbI$_3$ bi-layer interface.}
\label{BS-Interface}
\end{figure}

The bandgap values of either bulk or monolayer CsPbBr$_3$ (CsPbI$_3$) are attributed to the difference in energy level of Br-4p(I-5p) at the valence band maximum (VBM) and Pb-6p at the conduction band minimum (CBM) (Fig. S1, Fig. S2) \cite{Pitriana19}. Further, we have constructed a CsPbBr$_3$/CsPbI$_3$ bi-layer interface, where periodicity of atoms are there in the $y$-$z$ plane and vacuum of 15\AA \ is given in $x$-direction to understand the interface properties. In this case, the VBM is dominated by the I-5p than that of Br-4p, as shown in Fig. \ref{BS-Interface}(a,b). This dominance of I-5p over Br-4p is attributed to the lower electronegativity of I (2.66) that Br (2.96) \cite{Allred61}. At the interface, there are two types of iodine atoms are present. One at the interface, which is bonded to the Cs atom, and the other type is in the surface side. We elucidate that the 5p orbital of iodine atoms which are at the interface are responsible for valence band bending near the Fermi level more than 5p orbital of other atoms (Fig. \ref{BS-Interface}(c), Fig. S5). The cross sectional view of CsPbBr$_3$/CsPbI$_3$ bi-layer interface is shown in Fig. \ref{BS-Interface}(d).

\begin{figure}
\centering
\includegraphics[height=5.50cm, width=10.2cm]{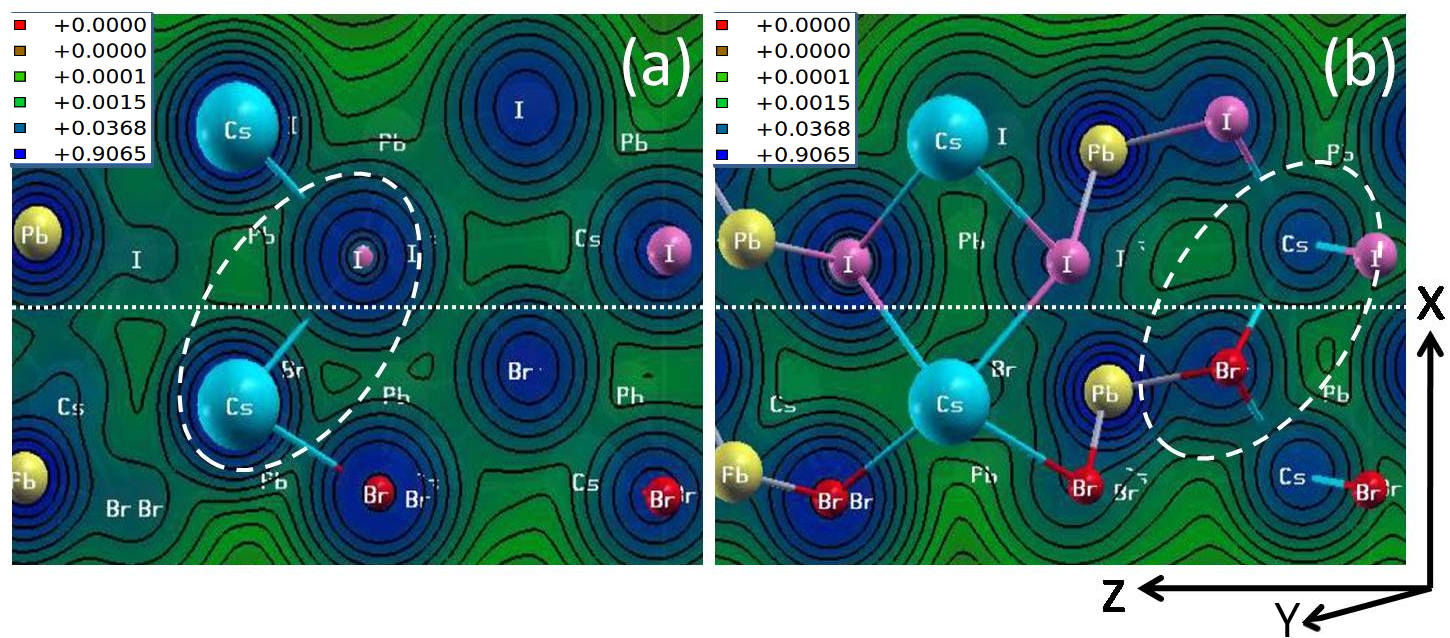}
\includegraphics[height=5.50cm, width=10.2cm]{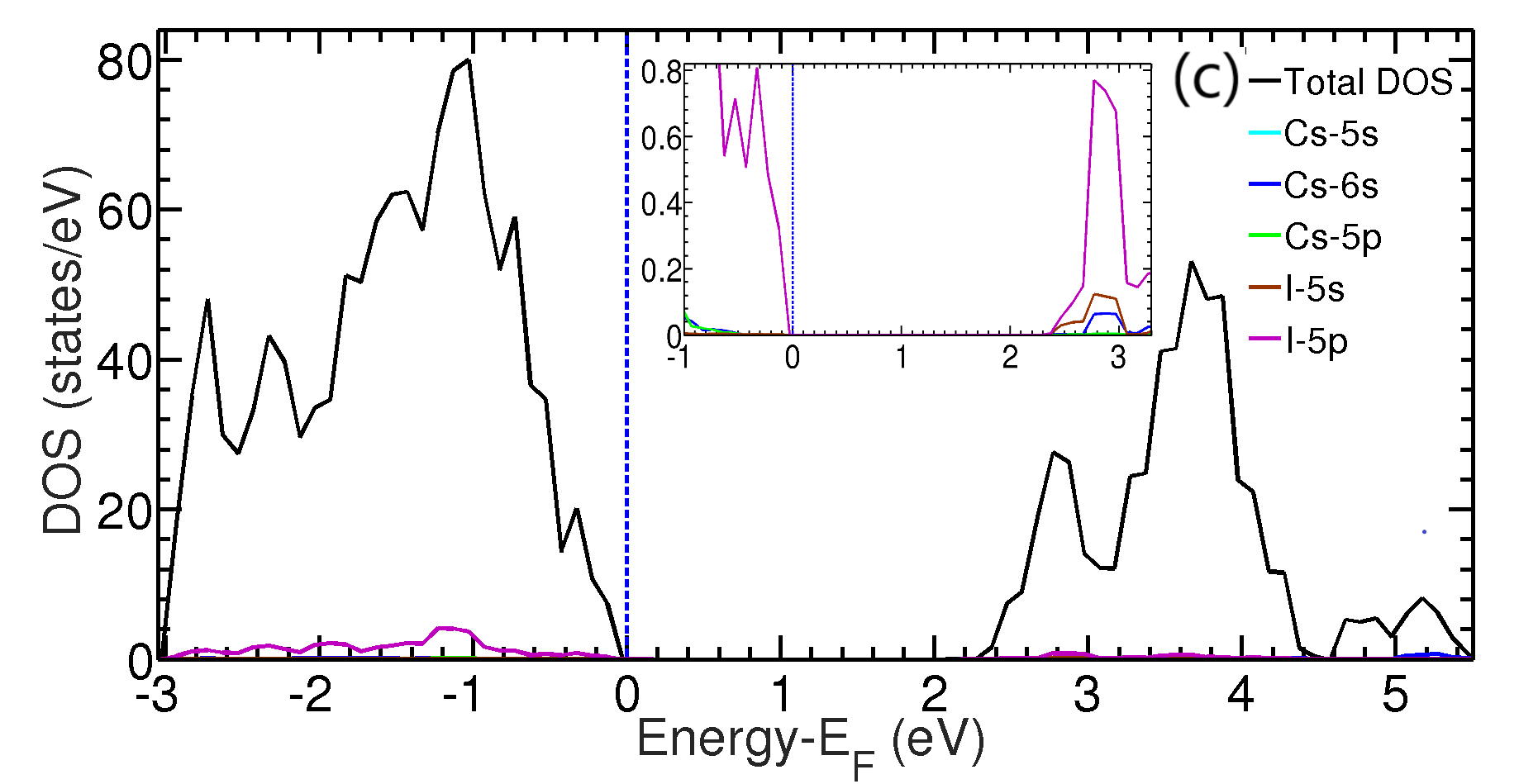}
\includegraphics[height=5.50cm, width=10.2cm]{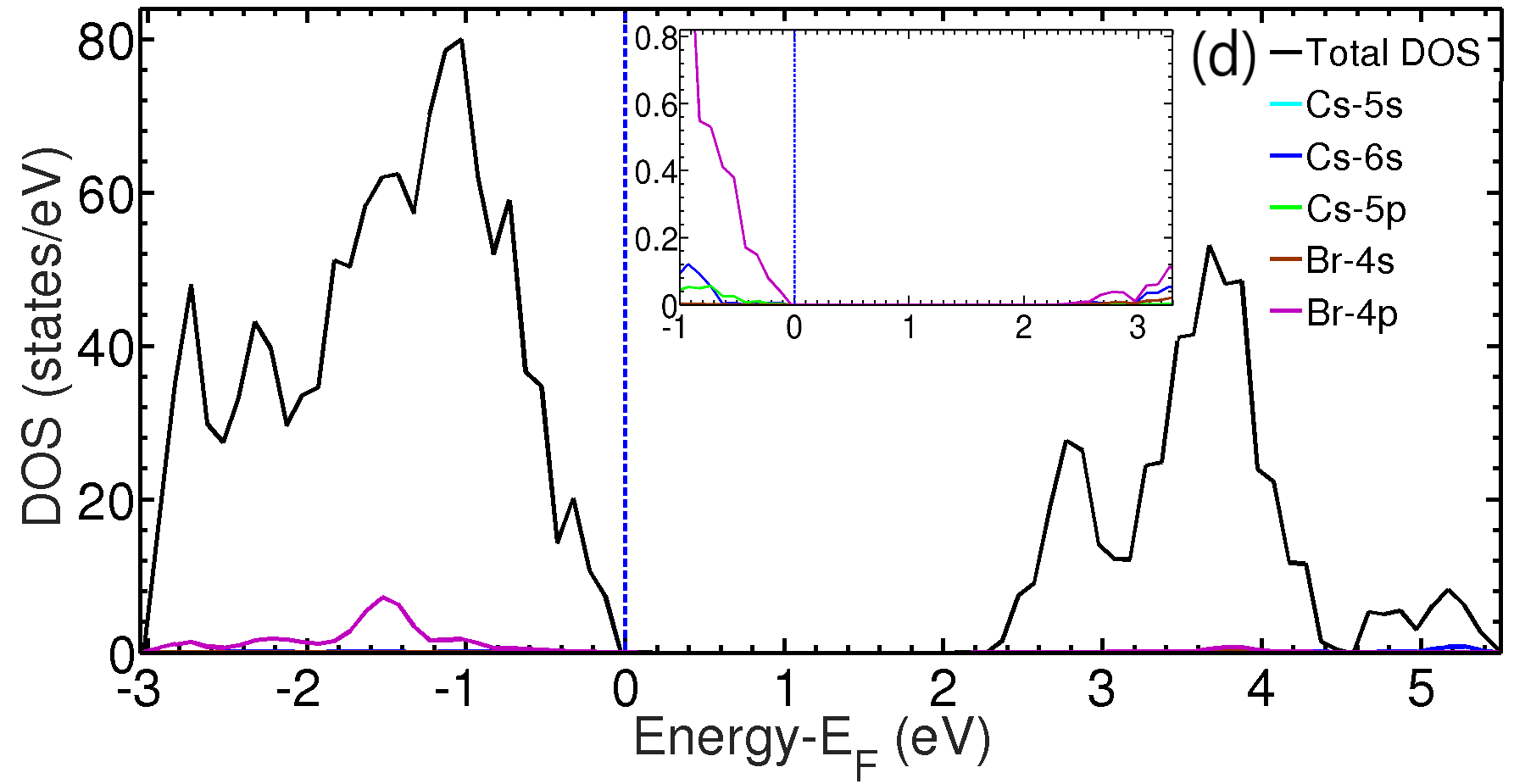}
\caption{Contour plot elucidating the charge density at the (a) Cs-I and (b) Cs-Br bonding at the CsPbBr$_3$/CsPbI$_3$ interface. Total density of states of CsPbBr$_3$/CsPbI$_3$ interface compared with partial electronic density of states of (c) Cs-I and (d) Cs-Br. Insets are zoomed versions of (c) and (d) near the Fermi level.}
\label{Charge-Density}
\end{figure}

The electron charge density at the interface is captured by the iso-electronic contour plot at a particular plane. We elucidate the two instances with Cs-I and Cs-Br bonding as shown in Fig. \ref{Charge-Density}(a) and Fig. \ref{Charge-Density}(b), respectively. It is revealed that more charge clouds are around the anion than of the cation indicating that the bonding is ionic in nature \cite{Pitriana19}. Fig. \ref{Charge-Density}(c) and \ref{Charge-Density}(d) show the projected density of state (PDOS) of the Cs-I and Cs-Br, contributing to the total DOS. Exact at the interface bonding environment, the contribution of Br-4p is much higher than that of any orbital contribution of Cs, both at the valence band and conduction band (inset in Fig. \ref{Charge-Density}(c)). Similar phenomenon is observed by I-5p for Cs-I bonding. The electronic band structure and PDOS of bulk CsPbBr$_3$ and CsPbI$_3$ are shown in Fig. S1. They reveal semiconducting characters with a band gap of 3.04 eV and 2.58 eV (in PBE level). In their monolayer structure, the band gap values are 3.88 eV and 3.15 eV respectively, owing to the quantum confinement. For the bi-layer structure, the band gap value is found out to be 2.76 eV. 

\makeatletter
\renewcommand{\fnum@figure}{Fig. \thefigure}

\makeatletter
\renewcommand{\fnum@figure}{Fig. \thefigure}
\makeatletter
\renewcommand{\fnum@figure}{Fig. \thefigure}
\begin{figure}
\centering
\includegraphics[height=6.5cm, width=10.2cm]{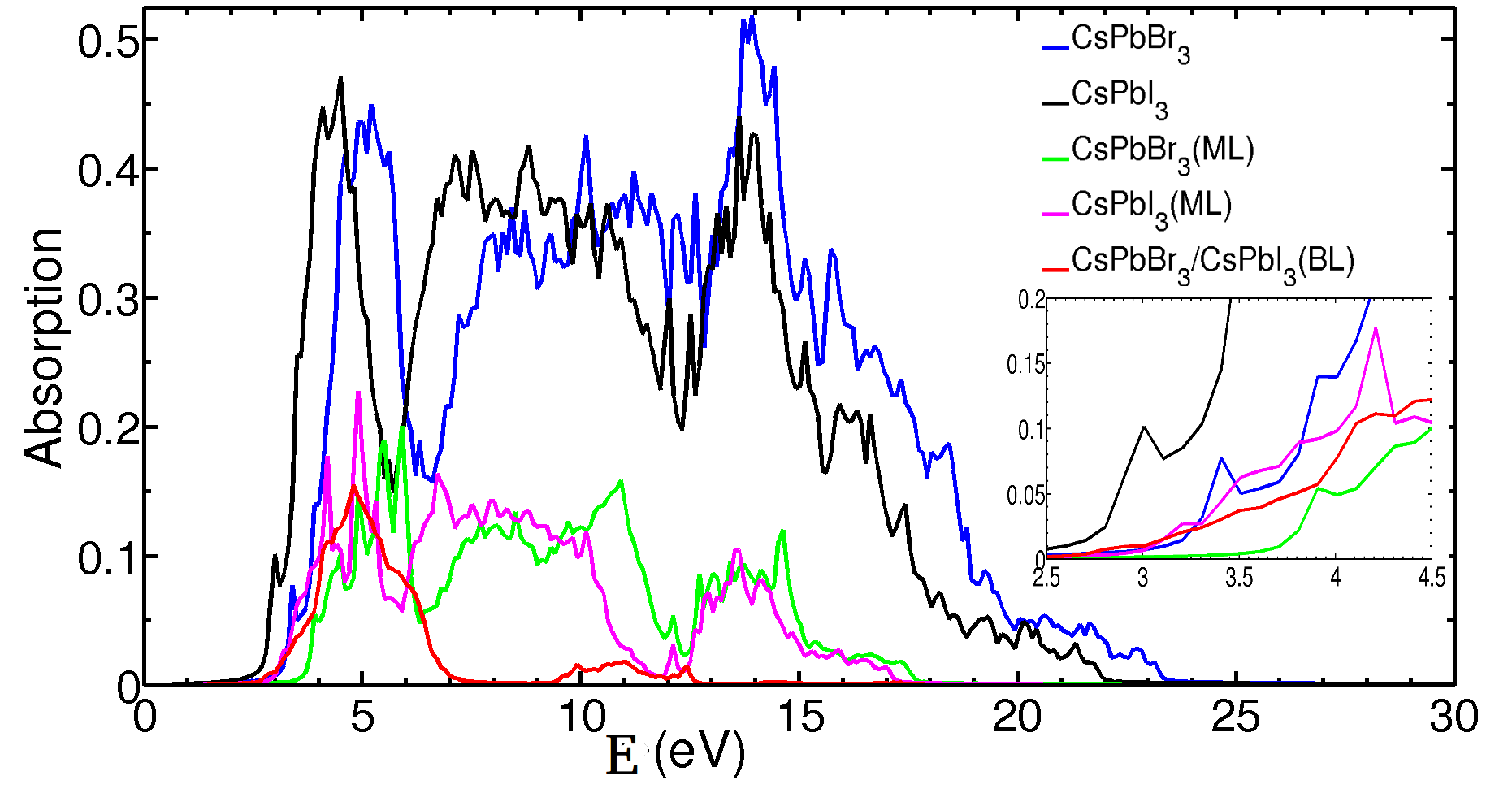}
\caption{Absorption spectra of CsPbBr$_{3-x}$I$_x$.}
\label{Absorption}
\end{figure}

The absorption spectrum of CsPbBr$_3$/CsPbI$_3$ interface geometry compared with their monolayer and bulk is shown in Fig. \ref{Absorption}. It is calculated by $\alpha (\omega)$ = $\sqrt{2} \omega_c \{[\varepsilon_1^2 (\omega) + \varepsilon_2^2 (\omega)]^{0.5} -\varepsilon_1 (\omega)\}^{0.5}$, where $\varepsilon_1$ and $\varepsilon_2$ are the real and imaginary part of the complex dielectric function, i.e. $\varepsilon (\omega)= \varepsilon_1(\omega) + i \varepsilon_2(\omega)$. It is observed that the reduced absorption intensity (for bilayer and monolayer geometries) is the consequence of geometrical confinement. The heavier halide reveals the red shift of the spectra both in bulk and confined geometries. At the lower range ($\sim$5 eV), the peak is due to halide Pb-6p, which is of low energy excitations. At the higher end ($\sim$14 eV), the peak is due to Pb-6s and Cs-6s, which reveal high energy excitations. In the intermediate range, the peaks are due to Cs-5p and Pb-6p. In the bilayer geometry, the lower energy peak is attributed to iodine more than bromine. However, the intermediate energy range peaks are heavily quenched and show blue shift, which is also observed in the superlattice geometry. We attribute the quenching of peaks to the shallow absorption probability.

The imaginary part of dielectric tensor $\varepsilon_2^{\alpha \beta}(\omega)$ is determined by $\varepsilon_2^{\alpha \beta}(\omega) = \dfrac{2\pi e^2}{\Omega \Eulerconst_0}+\sum\limits_{k,\nu,c} \delta(E^c_k-E^\nu_k-\hbar\omega)|<\psi^c_k|u.r|\psi^\nu_k>|^2$, where $\Eulerconst_0$ is the vacuum dielectric constant, $\Omega$ is the volume, $v$ and $c$ represent the valence and conduction bands, respectively, $\hbar\omega$ is the energy of the incident phonon, $u$ is the vector defining the polarization of the incident electric field, $u.r$ is the momentum operator, $\psi^c_k$ and $\psi^v_k$ are the wave functions of the conduction and valence bands at the $k$ point, respectively. The real part of the dielectric function can be written as $\varepsilon_1^{\alpha \beta}(\omega)$ = $1 +\dfrac{2}{\pi}P \int\limits_0^\infty \dfrac{\varepsilon_2^{\alpha \beta}(\omega') \omega'}{\omega'^2-\omega'^2+i\eta} d\omega'$, where $P$ is the principal value. The quenching of peak intensities at the intermediate to high energy range is also clearly observed at the imaginary and real part of the complex dielectric function for the bilayer geometry, unlike those of individual monolayer structures. 

\makeatletter
\renewcommand{\fnum@figure}{Fig. \thefigure}
\begin{figure}
\centering
\includegraphics[height=6.5cm, width=10.2cm]{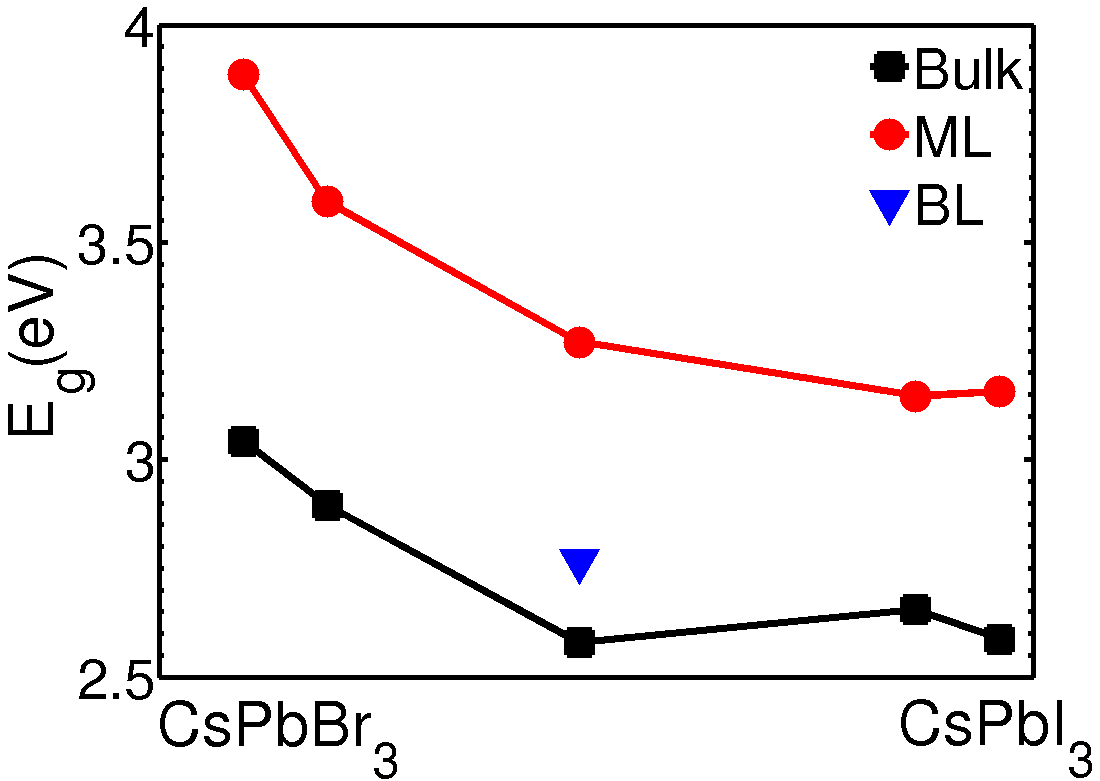}
\caption{Variation of bandgap with the increase of $x$ in CsPbBr$_{3-x}$I$_x$.}
\label{Band-gap}
\end{figure}
Figure \ref{Band-gap} shows the change in band gap with change in the Br/I content (\%) for bulk and monolayer cases. We point out the following features. (a) The change in band gap is more prominent for the substitution of bromine with iodine initially than later. (b) Oppositely, the change in band gap is less prominent for the substitution of iodine with bromine initially than later. (c) We attribute it to the presence of I-5p orbital near the Fermi level than that of Br-4p. (d) In the matrix of bromine (either in bulk or monolayer), a minute amount of iodine has more effect than vice versa.

Bulk CsPbBr$_3$ and CsPbI$_3$ possess orthorhombic crystal structures with space group Pnma. The optimized lattice parameters of bulk CsPbBr$_3$ are found to be a = 4.658 \AA, b = 10.025 \AA \ and c = 17.169 \AA \ and that of CsPbI$_3$ are found to be a = 4.903 \AA, b = 10.786 \AA \ and c = 18.215 \AA \ \cite{Ahmad17,Jain13}. With the change of halide, the change in structural parameter in orthorhombic phase is attributed to change in local octahedron distortion \cite{Chen18}. In bulk CsPbBr$_3$, the average Br-Pb bond distance is $\sim$3.078 \AA \ and that of average Br-I bond distance in CsPbI$_3$ is 3.282 \AA. In monolayer, it decreases to 8.8\% for CsPbBr$_3$ and 7.7\% for CsPbI$_3$. The decrease in bond lengths from bulk to monolayer is in accordance with the confinement in azimuthal direction.

For the bi-layer interface, Pb-Br and Pb-I stay at longer distance, hence do not make bonds (Fig. \ref{BS-Interface}(d)). However, the staggered bonding of Cs-Br and Cs-I with bond length of $\sim$ 4.106 and 3.922 \AA \ are revealed. Moreover, just above and below the interface planar, Pb-Br and Pb-I bonds are formed with bond lengths of 2.897 \AA \ and 3.087 \AA, \ respectively. 

\makeatletter
\renewcommand{\fnum@figure}{Fig. \thefigure}
\begin{figure}
\centering
\includegraphics[height=6.5cm, width=9.2cm]{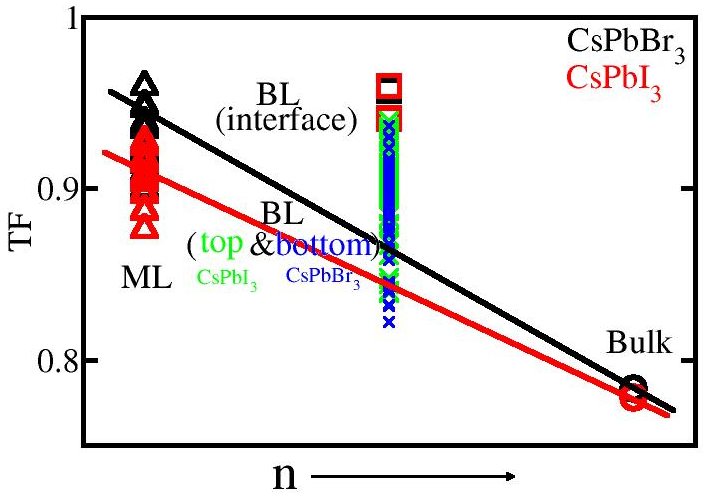}
\caption{Tolerance factor ($t$) of CsPbBr$_{3-x}$I$_x$ with number of layers ($n$).}
\label{Tolerance}
\end{figure}

The geometric ratio, called Goldschmidt's tolerance factor ($t$), assesses the stability of the A site cation within the cavities of BX$_3$ motif of the pervoskite structure. It is defined as $t= \dfrac{r_A+r_X}{\sqrt{2}(r_B+r_X)}$, where $r_A$ and $r_B$ are the ionic radii of the A and B site cations, respectively, and $r_X$ is the ionic radius of the anion \cite{Goldschmidt26}. With $t>1$, the formation of perovskite is highly improbable, whereas $t = 1$ indicates the perfect fit of the A site cation in BX$_3$ motif. The most favourable range of perovskite falls in $0.8 \leq t \leq 1$. When $t < 0.8$, A cation is too small, leading to distorted structure \cite{Tsui16,Travis16}.

\begin{figure}
\centering
\includegraphics[height=7.5cm, width=9.2cm]{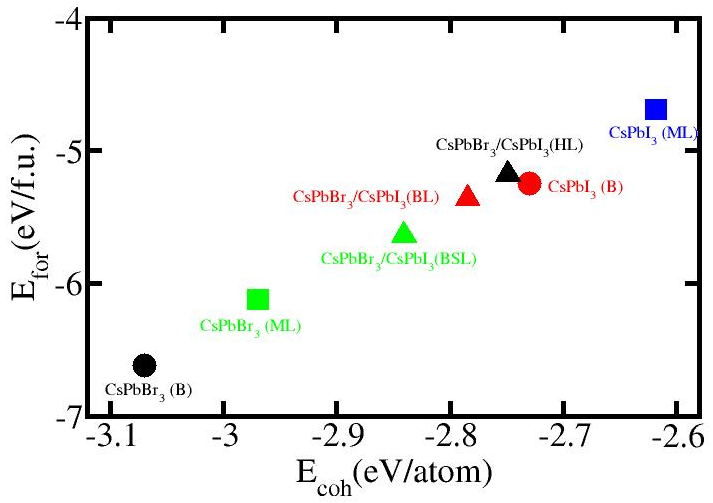}
\caption{Energy landscape of CsPbBr$_{3-x}$I$_x$.}
\label{Cohesive}
\end{figure}

Figure \ref{Tolerance} shows the distribution of $t$ for the bulk, bi-layer interface and monolayer structures. In bulk, $t < 0.8$ reflects the orthorhombic structures. For monolayer, $0.88 < t < 0.96$ indicates the partial distortions. It is seen that the value of $t$ is lower for CsPbI$_3$ than for CsPbBr$_3$ for both bulk and monolayer structures. However, for the bi-layer a wide range of values are observed. Careful observation leads to two groups, i.e. one at the interface and the other at the surface. A wider range of $t$ is observed at the surface than that of the interface. The value of $t$ at the surface is always lower than that of the interface. For CsPbBr$_3$ and CsPbI$_3$, the increase in $t$ is observed in order as $t_{\text{avg-bi-layer interface}} > t_{\text{avg-monolayer}} > t_{\text{bi-layer surface}} > t_{\text{bulk}}$. The shift of $t$ from $t < 0.8$ in the bulk to the range $0.8 < t < 1$ also reflects the structural rearrangement from anisotropy to isotropy in confinement.
Owing to the formulation, higher anionic dimension (I:198 pm and Br:185 pm) leads to lower value of $t$ (i.e., away from 1). Analyzing with formation energy (Fig. \ref{Cohesive}), compounds with bromine are more stable than that with iodine. Coincidentally, the lower electronegativity (I:2.66 and Br:2.96) of the heavier halides (I:126.904 u and Br:79.904 u) shows lower $t$. On the aspect of electronegartivity, substituting bromide with iodide leads to a narrowing of the bandgap. 

\makeatletter
\renewcommand{\fnum@figure}{Fig. \thefigure}

We investigate its energetic stability by calculating the formation energy (Supplementary data), which is found to be -6.61 eV for CsPbBr$_3$ and -5.24 eV for CsPbI$_3$. For monolayer CsPbBr$_3$ it is -6.12 eV and for monolayer CsPbI$_3$ it is -4.68 eV. We plot the energy landscape with formation energy and cohesive energy. Both for the bi-layer heterostructure and the superlattice heterostructures, the energies fall between their pristine components.

Interface formation energy can be calculated by the following formula 
 dE = (E$_{\text{CsPbBr}_3}$+ E$_{\text{CsPbI}_3}$ -E$_{\text{CsPbBr}_3}$/CsPbBr$_3$)/S, where E$_{\text{CsPbBr}_3}$ and E$_{\text{CsPbI}_3}$ are the total energy of individual monolayer and E$_{\text{CsPbBr}_3}$/CsPbBr$_3$ is total energy of the heterostructure, S is the area of the interface, respectively. The interface formation energy is -2.07 meV/\AA$^2$ and -7.22 meV/\AA$^2$ for the bi-layer heterostructure and superlattice heterostructure, respectively, which is comparable to other interface structures \cite{Bjorkman12,Zhang14}. We have examined the stability of the material by calculating the errors in total energy (<0.01Ry) and stress (<0.03GPa) due to change in kinetic energy cutoff and smearing parameter. The material is quite stable and the theoretical finding suggest that the interface is possible to form with suitable experimental conditions.

\section{Conclusion}
The structural and electronic properties of CsPbBr$_3$/CsPbI$_3$ are studied by density functional theory. The interface is formed by the staggered bonding of Cs-Br and Cs-I with bond length of $\sim$4.106 and 3.922 \AA. The wide range of tolerance factor for the bilayer geometry is observed, from which it is higher at the exact interface than at the surface. The change in band gap arises mainly due to the band bending of the I-5p orbital near the Fermi level than that of Br-4p. In the absorption spectra, the heavier halide shows the red shift for the pristine monolayer component, whereas for the bilayer geometry iodine contribution is mainly observed. The formation of bilayer structure is preferable over the iodine complex, but harder than the bromine complex. So, suitable synthesis procedures can be adopted for achieving the interface geometry for improved optical properties for better solar cell application.

\printcredits

\section*{Declaration of competing interest}
The authors declare that they have no known competing financial interests or personal relationships that could have appeared to influence the work reported in this paper. 

\section*{Acknowledgements}
We acknowledge the use of Xcrysden computer graphic tool and VESTA for visualizing the structures (www.xcrysden.org) \cite{Kokalj03, Moma11}.

\section*{Appendix A. Supplementary data}
Supplementary data related to this article can be found at https://doi.org/.

\bibliographystyle{cas-model2-names}

\bibliography{cas-refs}

\begin{thebibliography}{99}
\bibitem{Huang16} H. Huang, L. Polavarapu, J. A. Sichert, A. S. Susha, A. S. Urban, A. L. Rogach, NPG Asia Materials 8 (2016) e328.

\bibitem{Eperon17} G. E. Eperon, M. T. H\"orantner, H. J. Snaith, Nat. Rev. Chem. 1 (2017) 0095.

\bibitem{Duan19} J. Duan, H. Xu, W. E. I. Sha, Y. Zhao, Y. Wang, X. Yang, Q. Tang, J. Mater. Chem. A 7 (2019) 21036-21068.

\bibitem{Xiang19} W. Xiang, W. Tress, Adv. Mater. 31 (2019) 1902851.

\bibitem{Wang18} K. Wang, Z. Jin, L. Liang, H. Bian, D. Bai, H. Wang, J. Zhang, Q. Wang, S. Liu, Nat. Commun. 9 (2018) 4544.

\bibitem{Shu18} J. Shu, X. Zhang, P. Wang, R. Chen, H. Zhang, D. Li, P. Zhang, J. Xu, Physica B 548 (2018) 53-57. 


\bibitem{Gao19} Y. Gao, Y. Wu, H. Lu, C. Chen, Y. Liu, X. Bai, L. Yang, W. W. Yu, Q. Dai, Y. Zhang, Nano Energy 59 (2019) 517-526.

\bibitem{Protesescu15} L. Protesescu, S. Yakunin, M. I. Bodnarchuk, F. Krieg, R. Caputo, C. H. Hendon, R. X. Yang, A. Walsh, M. V. Kovalenko, Nano Lett. 15 (2015) 3692-3696.

\bibitem{Ramasamy16} P. Ramasamy, D. H. Lim, B. Kim, S. H. Lee, M. S. Lee, J. S. Lee, Chem. Commun. 52 (2016) 2067-2070.

\bibitem{Shen18} X. Shen, C. Sun, X. Bai, X. Zhang, Y. Wang, Y. Wang, H. Song, W.W. Yu, ACS Appl. Mater. Interfaces 10 (2018) 16768-16775.

\bibitem{Yantara15} N. Yantara, S. Bhaumik, F. Yan, D. Sabba, H. A. Dewi, N. Mathews, P. P. Boix, H. V. Demir, S. Mhaisalkar, J. Phys. Chem. Lett. 6 (2015) 4360-4364.

\bibitem{Bwang20} B. Wang, L. Liu, B. Liu, J. Li, B. Cao, Z. Zhao, Z. Liu, Physica B 599 (2020) 412488.


\bibitem{Fadla20} M. A. Fadla, B. Bentria, T. Dahame, A. Benghia, Physica B 585 (2020) 412118.

\bibitem{Yoon16} H. C. Yoon, H. Kang, S. Lee, J. H. Oh, H. Yang, Y. R. Do, ACS Appl. Mater. Interfaces 8 (2016), 18189-18200.

\bibitem{Guan20} H. Guan, S. Zhao, H. Wang, D. Yan, M. Wang, Z. Zang, Nano Energy 67 (2020) 104279.

\bibitem{Yan21} D. Yan, S. Zhao, Y. Zhang, H. Wang, Z. Zang, Opto-Electron Adv. 4 (2021) 200075.

\bibitem{Sidhik17} S. Sidhik, D. Esparza, A. Mart\'inez-Ben\'itez, T. Lopez-Luke, R. Carriles, I. Mora-Sero, E. de la Rosa, J. Phys. Chem. C 121 (2017) 4239-4245.

\bibitem{Carella18} A. Carella, R. Centore, F. Borbone, M. Toscanesi, M. Trifuoggi, F. Bella, C. Gerbaldi, S. Galliano, E. Schiavo, A. Massaro, A. B. Mu$\tilde{\text{n}}$oz-Garc{\'i}a, M. Pavone, Electrochimica Acta, 292 (2018) 805-816.

\bibitem{Fagiolari20} L. Fagiolari, M. Bonomo, A. Cognetti, G. Meligrana, C. Gerbaldi, C. Barolo, F. Bella, ChemSusChem 13 (2020) 6562.

\bibitem{Rahman21} N. A. Rahman, S. A. Hanifah, N. N. Mobarak, A. Ahmad, N. A. Ludin, F. Bella, M. S. Su'ait, Polymer 230 (2021) 124092.

\bibitem{Haro21} J. C. de Haro, E. Tatsi, L. Fagiolari, M. Bonomo, C. Barolo, S. Turri, F. Bella, G. Griffini, ACS Sustainable Chem. Eng. 9 (2021) 8550-8560.

\bibitem{Yang21} K. Yang, S. Liu, J. Du, W. Zhang, Q. Huang, W. Zhang, W. Hu, Y. Hu, Y. Rong, A. Mei, H. Han, Sol. RRL. 5 (2021) 2000825.

\bibitem{Lu22} C. Lu, M. Aftabuzzaman, C. H. Kim, H. K. Kim, Chem. Engig. J. 428 (2022) 131108.

\bibitem{Dokouzis20} A. Dokouzis, F. Bella, K. Theodosiou, C. Gerbaldi, G. Leftheriotis, Mater. Today Energy 15 (2020) 100365.

\bibitem{Parikh22} N. Parikh, M. Karamta, N. Yadav, M. M. Tavakoli, D. Prochowicz, S. Akin, A. Kalam, S. Satapathi, P. Yadav, J. Ene. Chem. 66 (2022) 4-90.

\bibitem{Kumar21} N. Kumar, J. Rani, R. Kurchania, Solar Energy 221 (2021) 197-205.

\bibitem{Ding21} X. Ding, Y. Zhang, F. Sheng, Y. Li, L. Zhi, X. Cao, X. Cui, D. Zhuang, J. Wei, ACS Appl. Energy Mater. 4 (2021) 5504-5510.

\bibitem{Pitriana19} P. Pitriana, T. D. K. Wungu, Hermana, R. Hidayat, Results in Physics 15 (2019) 102592.

\bibitem{Nam17} J. K. Nam, S. U. Chai, W. Cha, Y. J. Choi, W. Kim, M. S. Jung, J. Kwon, D. Kim, J. H. Park, Nano Lett. 17 (2017) 2028-2033.

\bibitem{Li18} Y. Li, J. Duan, H. Yuan, Y. Zhao, B. He, Q. Tang, Solar RRL 2 (2018) 1800164.

\bibitem{Akkerman17} Q. A. Akkerman, D. Meggiolaro, Z. Dang, F. D. Angelis, L. Manna, ACS Energy Lett. 2 (2017) 2183-2186.

\bibitem{Hu17} Y. Hu, F. Bai, X. Liu, Q. Ji, X. Miao, T. Qiu, S. Zhang, ACS Energy Lett. 2 (2017) 2219-2227.

\bibitem{Liang17} J. Liang, P. Zhao, C. Wang, Y. Wang, Y. Hu, G. Zhu, L. Ma, J. Liu, Z. Jin, J. Am. Chem. Soc. 139 (2017) 14009-14012.

\bibitem{Goesten18} M. G. Goesten, R. Hoffmann, J. Am. Chem. Soc. 140 (2018) 12996-13010.

\bibitem{Bian18} H. Bian, D. Bai, Z. Jin, K. Wang, L. Liang, H. Wang, J. Zhang, Q. Wang, S. Liu, Joule 2 (2018) 1500-1510.

\bibitem{Chen18} X. Chen, D. Han, Y. Su, Q. Zeng, L. Liu, D. Shen, Phys. Stat. Sol. Rap. Res. Lett. 12 (2018) 1800193.

\bibitem{Haque20} A. Haque, T. D. Chonamada, A. B. Dey, P. K. Santra, Nanoscale 12 (2020) 20840-20848.

\bibitem{Murtaza11} G. Murtaza, I. Ahmad, Physica B 406 (2011) 3222-3229.

\bibitem{Yan20} L. Yan, M. Wang, C. Zhai, L. Zhao, S. Lin, ACS Applied Materials \& Interfaces 12 (2020) 40453-40464.

\bibitem{Ghaithan20} H. M. Ghaithan, Z. A. Alahmed, A. Lyras, S. M. H. Qaid, A. S. Aldwayyan, Crystals 10 (2020) 342. 

\bibitem{Yaniv78} A. Yaniv, Phy. Rev. B 17 (1978) 3904-3918.

\bibitem{Oliver12} D. J. Oliver, J. Maassen, M. E. Ouali, W. Paul, T. Hagedorn, Y. Miyahara, Y. Qi, H. Guo, P. Gr{\"u}tter, Proc. Nat. Aca. Sci, USA, 109 (2012) 19097-19102.

\bibitem{Saha11} B. Saha, T. D. Sands, U. V. Waghmare, J. Appl. Phys. 109 (2011) 083717.

\bibitem{Samal11} D. Samal, P. S. Anil Kumar, J. Supercond. Nov. Magn. 24 (2011) 915-918.

\bibitem{Wang20} H. Wang, F. Tang, P. H. Dhuvad, X. Wu, npj Comput. Mater. 6 (2020) 52.

\bibitem{Chikina21} A. Chikina, D. V. Christensen, V. Borisov, M.-A. Husanu, Y. Chen, X. Wang, T. Schmitt, M. Radovic, N. Nagaosa, A. S. Mishchenko, R. Valent\'i, N. Pryds, V. N. Strocov, ACS Nano 15 (2021) 4347-4356.

\bibitem{Eshkalak18} K. E. Eshkalak, S. Sadeghzadeh, M. Jalaly, Solid State Comm. 270 (2018) 82-86.

\bibitem{Arora21} A. Arora, P. K. Nayak, S. Bhattacharyya, N. Maity, A. K. Singh, A. Krishnan, M. S. R. Rao, Phy. Rev. B 103 (2021) 205406.

\bibitem{Kosmider13} K. Kosmider, J. Fernandez-Rossier, Phy. Rev. B 87 (2013) 075451.

\bibitem{Li10} B. -W. Li, M. Osada, T. C. Ozawa, Y. Ebina, K. Akatsuka, R. Ma, H. Funakubo, T. Sasaki, ACS Nano 4 (2010) 6673-6680.

\bibitem{Yamada02} H. Yamada, M. Kawasaki, Y. Ogawa, Y. Tokura, Appl. Phys. Lett. 81 (2002) 4793.

\bibitem{Tsukada97} I. Tsukada, M. Nose, K. Uchinokura, Physica C 282-287 (1997) 687-688.

\bibitem{Giannozzi09} P. Giannozzi, P. Giannozziand, S. Baroni, N. Bonini, M. Calandra, R. Car, C.Cavazzoni, D. Ceresoli, G.L. Chiarotti, M. Cococcioni, I. Dabo, A. D. Corso, S. Fabris, G. Fratesi, S. de Gironcoli, R. Gebauer, U. Gerstmann, C. Gougoussis, A. Kokalj, M. Lazzeri, L. Martin-Samos, N. Marzari, F. Mauri, R. Mazzarello, S. Paolini, A. Pasquarello, L. Paulatto, C. Sbraccia, S. Scandolo, G. Sclauzero, A.P. Seitsonen, A. Smogunov, P. Umari and R. M. Wentzcovitch, J. Phys. Condens. Matter. 21 (2009) 395502.

\bibitem{Perdew96} J. P. Perdew, K. Burke, M. Ernzerhof, Phys. Rev. Lett. 77 (1996) 3865-3868.

\bibitem{Vanderbilt90} D. Vanderbilt, Phys. Rev. B 41 (1990) 7892-7895. 

\bibitem{Monkhorst76} H. J. Monkhorst, J. D. Pack, Phys. Rev. B 13 (1976) 5188-5192. 

\bibitem{Methfessel89} M. Methfessel, A. Paxton, Phys. Rev. B 40 (1989) 3616-3621.

\bibitem{Allred61} A. L. Allred, J. Inorg. Nucl. Chem. 17 (1961) 215-221.

\bibitem{Ahmad17} M. Ahmad, G. Rehman, L. Ali, M. Shafiq, R. Iqbal, R. Ahmad, T. Khan, S. Jalali-Asadabadi, M. Maqbool, I. Ahmad, J. Alloys and Comp. 705 (2017) 828-839.

\bibitem{Jain13} A. Jain, S. P. Ong, G. Hautier, W. Chen, W. D. Richards, S. Dacek, S. Cholia, D. Gunter, D. Skinner, G. Ceder, K. A. Persson, APL Materials 1 (2013) 011002.

\bibitem{Goldschmidt26} V. M. Goldschmidt, Naturwissenschaften, 14 (1926) 477-485. 

\bibitem{Tsui16} K. Y. Tsui, N. Onishi, R. F. Berger, J. Phys. Chem. C 120 (2016) 23293-23298.

\bibitem{Travis16} W. Travis, E. N. K. Glover, H. Bronstein, D. O. Scanlon, R. G. Palgrave, Chem. Sci. 7 (2016) 4548-4556.

\bibitem{Bjorkman12} T. Bjorkman, A. Gulans, A.V. Krasheninnikov, R. M. Nieminen, Phys. Rev. Lett. 108 (2012) 235502.

\bibitem{Zhang14} H. Zhang, Y.-N. Zhang, H. Liu, L.-M. Liu, J. Mater. Chem. A 2 (2014) 15389-15395.

\bibitem{Kokalj03} A. Kokalj, Comp. Mater. Sci. 28 (2003) 155-168.

\bibitem{Moma11} K. Momma, F. Izumi, J. Appl. Crystallogr. 44 (2011) 1272-1276.

\end{thebibliography}

\end{document}